\documentclass[aip,reprint,jcp]{revtex4-1}

\usepackage{graphicx}
\usepackage{dcolumn}
\usepackage{epstopdf}
\usepackage{hyperref}

\begin{document}

\title{Entropy based fingerprint for local crystalline order}

\author{Pablo M. Piaggi}
\affiliation{Theory and Simulation of Materials (THEOS), {\'E}cole Polytechnique F{\'e}d{\'e}rale de Lausanne, c/o USI Campus, Via Giuseppe Buffi 13, CH-6900, Lugano, Switzerland}
\affiliation{Facolt{\`a} di Informatica, Instituto di Scienze Computazionali, and National Center for Computational Design and Discovery of Novel Materials MARVEL, Universit{\`a} della Svizzera italiana (USI), Via Giuseppe Buffi 13, CH-6900, Lugano, Switzerland}
\author{Michele Parrinello}%
\email{parrinello@phys.chem.ethz.ch}
\affiliation{Department of Chemistry and Applied Biosciences, ETH Zurich, c/o USI Campus, Via Giuseppe Buffi 13, CH-6900, Lugano, Switzerland}
\affiliation{Facolt{\`a} di Informatica, Instituto di Scienze Computazionali, and National Center for Computational Design and Discovery of Novel Materials MARVEL, Universit{\`a} della Svizzera italiana (USI), Via Giuseppe Buffi 13, CH-6900, Lugano, Switzerland}

\date{\today}

\begin{abstract}
  We introduce a new fingerprint that allows distinguishing between liquid-like and solid-like atomic environments.
  This fingerprint is based on an approximate expression for the entropy projected on individual atoms.
  When combined with a local enthalpy, this fingerprint acquires an even finer resolution and it is capable of discriminating between different crystal structures.
\end{abstract}

\maketitle

\section{Introduction}

Atomistic computer simulation is an important technique used in the study of a broad range of phenomena in materials science, chemistry, and condensed matter physics.
In these fields, very often one is faced with the problem of identifying different local arrangements.
A paradigmatic case is that of the nucleation of a crystal from the liquid where one is required to distinguish between solid-like and liquid-like atomic environments.
The situation is even more complicated in systems exhibiting polymorphism since in these cases it is desirable to classify the atoms as belonging to one of the different polymorphic structures.
This is a common occurrence in nucleation studies where Ostwald's step rule is observed \cite{tenWolde99,Giberti15} or where clusters exhibit a core-shell structure \cite{tenWolde95,Lechner11}.
Another area where the ability to distinguish between different local arrangements plays a role is in the identification of crystallites in nanocrystalline materials \cite{Meyers06}.

Several methods have been proposed to distinguish between liquid-like and solid-like atoms and to identify local crystalline structures.
One such method is the common neighbor analysis (CNA) \cite{Honeycutt87,Stukowski12} which is an efficient algorithm able to distinguish between liquid, bcc, fcc, and hcp phases.
However, it lacks robustness with respect to particle displacements such as those arrising from thermal motion or stresses.
Another popular method is based on the local Steinhardt parameters \cite{Lechner08} which are local, averaged versions of the original Steinhardt parameters \cite{Steinhardt83}.
However, they also come at a high computational cost and presume that the nature of the crystal structure is known beforehand.

This work is inspired by a recent progress in the study of nucleation using metadynamics\cite{Laio02,Barducci08} to enhance the probability of inducing the crystal formation in an accessible computer time.
Metadynamics relies on the identification of appropriate collective variables (CVs).
In Ref.~\citenum{Piaggi17} we found that enthalpy and an approximate expression for entropy based on the two body correlation function, were useful CVs in this context.
One of the features of this work was that the CVs did not contain any information on the geometry of the crystal structure.
This suggested that perhaps from these two quantities one could extract fingerprints able to distinguish between different local atomic arrangements.

Enthalpy and entropy are global properties and in order to be able to use them as local parameters we have to project them onto each atom.
We propose a method that is able to do so.
We find that the local entropy thus defined is able to distinguish extremely well between solid-like and liquid-like atoms.
Furthermore, in conjuction with local enthalpy it can distinguish well between different polymorphs, even in the subtle case of the difference between fcc-like and hcp-like arrangements.

\section{Entropy approximation based on the two body correlation function}

\begin{figure*}
	\begin{center}
		\includegraphics[width=0.95\textwidth]{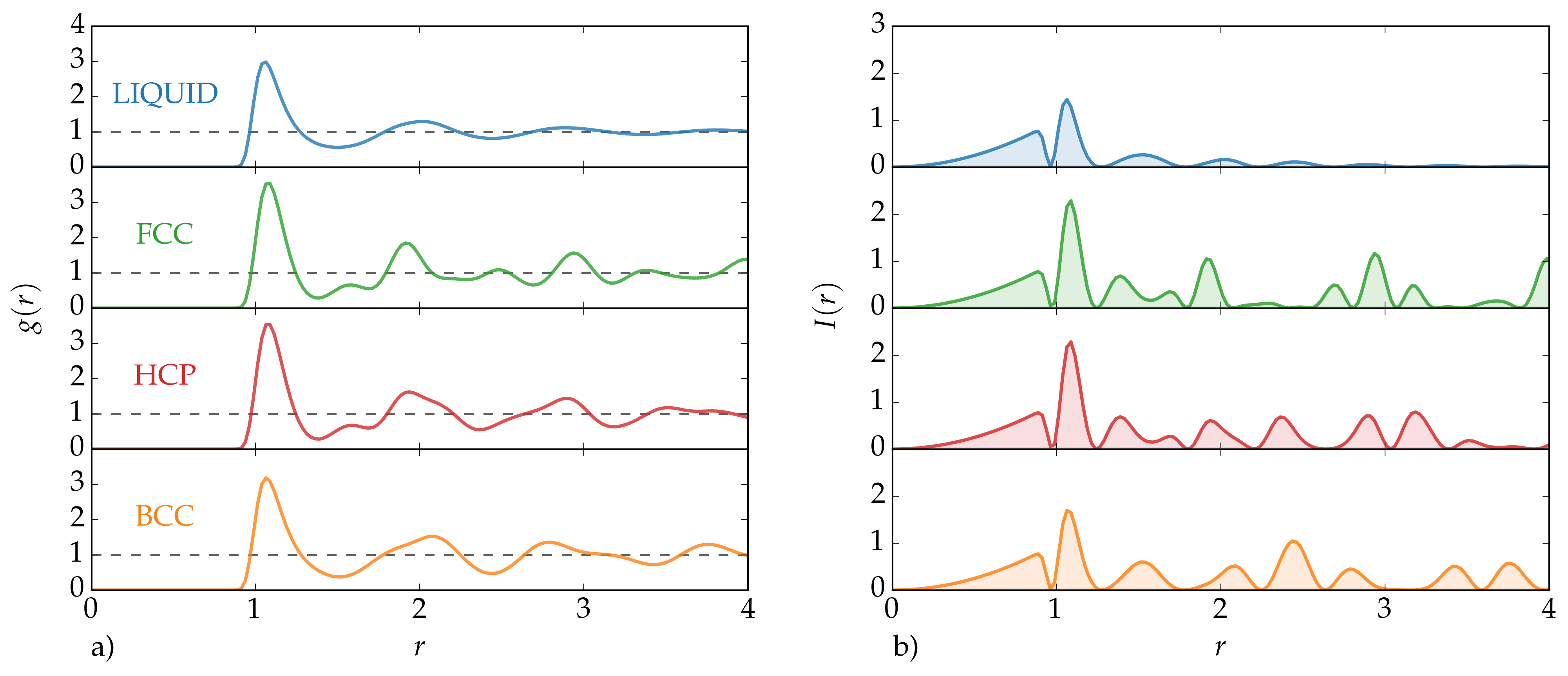}
		\caption{\label{fig:Figure1}
    Analysis of functions related to the entropy approximation $S_2$.
    a) Radial distribution function $g(r)$, and b) integrand in Eq.~(\ref{eq:pair_entropy}) $I(r)=[ g(r) \ln g(r) - g(r) + 1 ] r^2$.
    These functions are compared for the liquid, fcc, hcp, and bcc phases of a Lennard-Jones fluid at the melting temperature.
    We use Lennard-Jones units, i.e. $\sigma=1$.
    }%
	\end{center}
\end{figure*}

\label{sec:PairEntropy}
Ref.~\citenum{Piaggi17} was based on the consideration that in the liquid to solid transition there is a trade-off between entropy and enthalpy.
The role of metadynamics was there to enhance the fluctuations of these two quantities so as to accelerate crystallization.
This required designing CVs able to describe these two quantities.
Enthalpy is easy to compute but entropy is extremely costly to evaluate.
However, an expression that gives an approximate evaluation of the entropy is sufficient for the purpose of driving crystallization.
Such an expression was derived from an expansion of the configurational entropy in terms of multibody correlation functions\cite{Green52,Nettleton58,Baranyai89}.
In simple liquids the second term of the expansion, often called two-body excess entropy, involves only the pair correlation function and accounts for about 90\% of the configurational entropy \cite{Wallace87,Wallace94,Laird92,Baranyai89}.
This term is given by,
\begin{equation}
  S_2 = -2\pi\rho k_B \int\limits_0^{\infty} \left [ g(r) \ln g(r) - g(r) + 1 \right ] r^2 dr,
  \label{eq:pair_entropy}
\end{equation}
where $\rho$ is the system's density, and $g(r)$ is the radial distribution function.
Extensions of the expansion to multicomponent\cite{Hernando90,Prestipino04} and inhomogeneus\cite{Morita61} systems are also available.
We also recall that entropy series expansions have been used to study order-disorder phenomena starting with the landmark work of Kikuchi\cite{Kikuchi51}.

In order to come to grasp with $S_2$ and understand better why it works, we first contrast in Fig. \ref{fig:Figure1} the different behaviors of $g(r)$ and the integral in Eq.~\ref{eq:pair_entropy} $I(r)=[ g(r) \ln g(r) - g(r) + 1 ] r^2$.
The data were taken from a system with Lennard-Jones interactions at temperature $T=1.15$ and pressure $P=5.68$, that corresponds to the solid-liquid coexistence point\cite{Hansen69}.
The Lennard-Jones potential was truncated at 2.5 and tail corrections were included.
We refer the reader to Appendix \ref{sec:appendixA} for further computational details.
As usual we use Lennard-Jones units \cite{FrenkelBook}, i.e. $\sigma=1$ and $\epsilon=1$.
We have chosen these thermodynamic conditions because at this temperature and pressure the fcc, hcp, bcc, and liquid phases are all metastable allowing a fair comparison.
The first observation is that while $g(r)$ has some difficulty at distinguishing betweem solid and liquid, it strikes the eye that $I(r)$ in the liquid phase is much more short ranged than in the solid phases.
Furthermore, the $g(r)$ for the solid phases can hardly distinguish between the different polymorphs.
In contrast, the bcc $I(r)$ appears clearly different from that of the closed packed structures.
More subtle is the difference between fcc and hcp, that is revealed only if one goes as far out as the third neighbor shell.

\section{Entropy fingerprint for solid-like and liquid-like environments}

\label{sec:entropy_fingerprint}

The analysis of $I(r)$ suggests that, if properly projected onto the different atoms, $S_2$ could be used as a fingerprint to identify local structures.
The projection on atom $i$ can be achieved using the expression:
\begin{equation}
  s_S^i = -2\pi\rho k_B \int\limits_0^{r_{m}} \left [ g_m^i(r) \ln g_m^i(r) - g_m^i(r) + 1 \right ] r^2 dr,
  \label{eq:entropy_parameter}
\end{equation}
where $r_{m}$ is an upper integration limit that in principle should be taken to infinity, and $g_m^i$ is the radial distribution function centered at the $i$-th atom.
To obtain a continuous and differentiable order parameter, we define a mollified version of the radial distribution function\cite{Piaggi17},
\begin{equation}
        g_m^i(r) = \frac{1}{4 \pi \rho r^2} \sum\limits_{j} \frac{1}{\sqrt{2 \pi \sigma^2}} e^{-(r-r_{ij})^2/(2\sigma^2)} ,
  \label{eq:mollified_rdf}
\end{equation}
where $j$ are the neighbors of atom $i$, $r_{ij}$ is the distance between atoms $i$ and $j$, and $\sigma$ is a broadening parameter.
We shall choose $\sigma$ so small that $g_m(r) \sim g(r)$ yet large enough for the derivatives relative to the atomic positions to be manageable\cite{Piaggi17}.
A similar projection of $S_2$ has been used in Ref. \citenum{Leocmach13}.

If we use $s_S^i$ as defined in Eq. (\ref{eq:entropy_parameter}) it can be seen in Fig. \ref{fig:Figure2} that, in the cases of Na\cite{Wilson15} at 350 K and Al\cite{Sturgeon00} at 900 K (see Appendix \ref{sec:appendixA} for technical details), the distribution of $s_S^i$ in the liquid and solid phases are peaked at two different positions but exhibit a large overlap.
\begin{figure*}
	\begin{center}
		\includegraphics[width=0.95\textwidth]{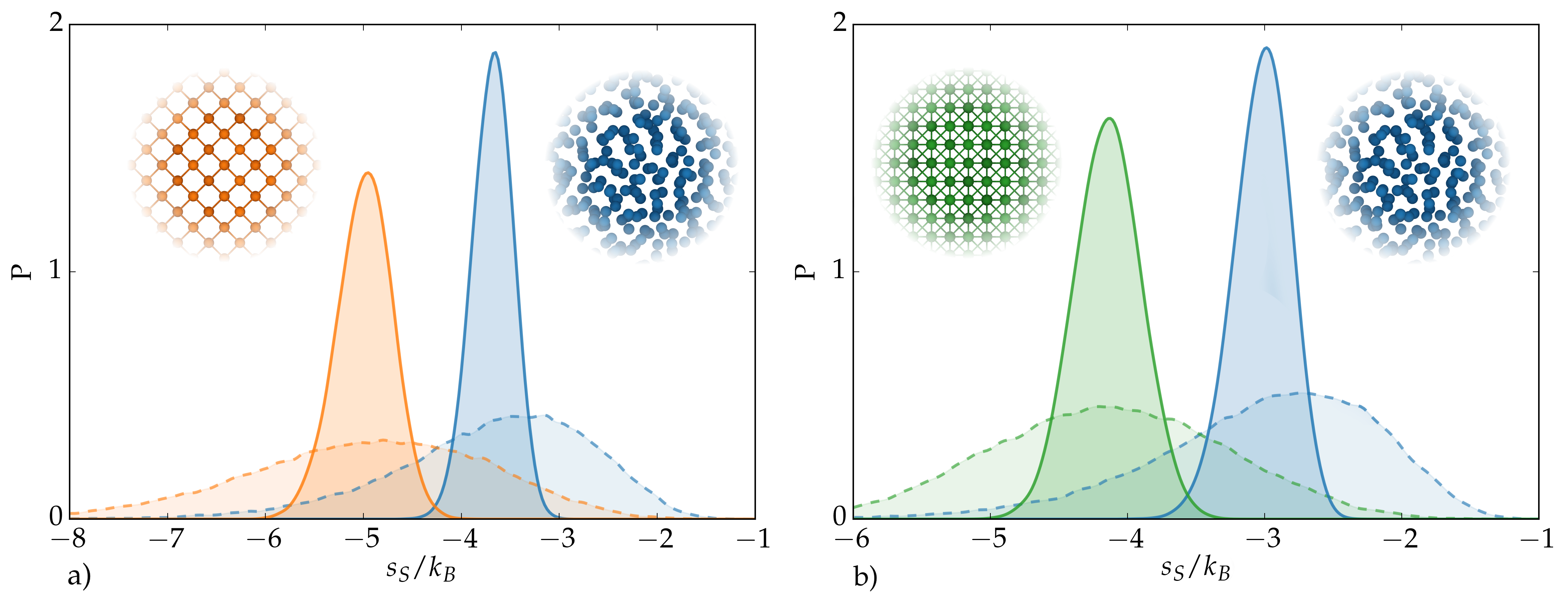}
		\caption{\label{fig:Figure2} Distributions of $s_S$ and $\bar{s}_S$ for a) bcc Na\cite{Wilson15} at 350 K, b) fcc Al\cite{Sturgeon00} at 900 K.
    Orange, green, and blue lines refer to the bcc, fcc, and liquid phases, respectively.
    Dotted and full lines refer to the non-averaged $s_S$ and averaged parameters $\bar{s}_S$, respectively.
    The probability distributions of each phase are normalized to one.
    The solid atomic configurations correspond to $\{100\}$ planes of bcc and fcc crystals at 0 K.
    The parameters $r_{m}$, $r_{a}$, and $\sigma$ that were used are summarized in Table \ref{tab:table1}.
    }
	\end{center}
\end{figure*}
In order to calculate local order parameters whose distributions are more clearly distinct, we take cue from Lechner and Dellago \cite{Lechner08} and define an average local entropy:
\begin{equation}
  \bar{s}_S^i = \frac{\sum_j s_S^j f(r_{ij}) + s_S^i}{\sum_j f(r_{ij})+1}
  \label{eq:entropy_parameter_avg}
\end{equation}
where $j$ runs over the neighbors of atom $i$ and $f(r_{ij})$ is a switching function with cutoff $r_{a}$.
Switching functions have a value of 1 for $r_{ij} \ll r_{a}$, 0 for $r_{ij} \gg r_{a}$, and decay smoothly from 1 to 0 for $r_{ij} \approx r_{a}$.
We have used a switching function with the functional form:
\begin{equation}
  f(r_{ij}) = \frac{1-(r_{ij}/r_{a})^N}{1-(r_{ij}/r_{a})^M}
\end{equation}
with $N=6$ and $M=12$.
Such a form has proven useful in many other contexts\cite{Tribello14}.
At variance with $s_S^i$, the distributions of $\bar{s}_S^i$ of the liquid and solid phases now have a negligible overlap (see Fig. \ref{fig:Figure2}).
Henceforth, we shall drop the index $i$ when referring to distributions and we shall refer to $\bar{s}_S$ as entropy fingerprint.

The ability to distinguish sharply between solid-like and liquid-like molecules depend on a wise choice of the parameters $r_m$ and $r_a$.
As $r_m$ is increased, more of the long range part of the integrand is included making the difference between liquid and solid more and more evident.
On the other hand by increasing $r_a$, more neighbors are included in the summation in Eq. (\ref{eq:entropy_parameter_avg}) and eventually the locality of $\bar{s}_S$ is lost.
In the practice we have chosen for $r_m$ and $r_a$ the smallest values that still ensure sharp distinction between solid-like and liquid-like atoms.
The parameters $r_{m}$, $r_{a}$, and $\sigma$ that were used are summarized in Table \ref{tab:table1}.
\begin{table}[b]
\caption{\label{tab:table1} Parameters in the definition of $s_S$ and $\bar{s}_S$ for different structures.
  The columns represent the crystal structure, the model system, the temperature (T) at which the distributions of solid and liquid phases are compared, and the parameters $r_{m}$ and $r_{a}$ defined in Eq.~(\ref{eq:entropy_parameter}) and (\ref{eq:mollified_rdf}).
  $r_{m}$ and $r_{a}$ are in units of the lattice constant, $a=4.23$ \AA~for Na and $a=4.05$ \AA~for Al.
  We report the number of neighbor shells (NS) corresponding to $r_m$ and $r_a$.
  For both cases $\sigma=0.02$ nm.
  }
\begin{ruledtabular}
\begin{tabular}{ccccc}
Structure & Model & T (K) & $r_{m}$ ($a$) & $r_{a}$ ($a$) \\
\hline
bcc & Na & 350 & 1.8 (5NS) & 1.2 (2NS) \\
fcc & Al & 900 & 1.4 (3NS) & 0.9 (1NS) \\
\end{tabular}
\end{ruledtabular}
\end{table}

It is interesting to investigate whether the entropy fingerprint can identify ordered structures in a complex situation, in a context different from nucleation.
To this effect we generated a nanocrystalline structure (see Fig. \ref{fig:Figure3}) using a procedure described in Appendix \ref{sec:appendixA}.
\begin{figure}
	\begin{center}
		\includegraphics[width=0.48\textwidth]{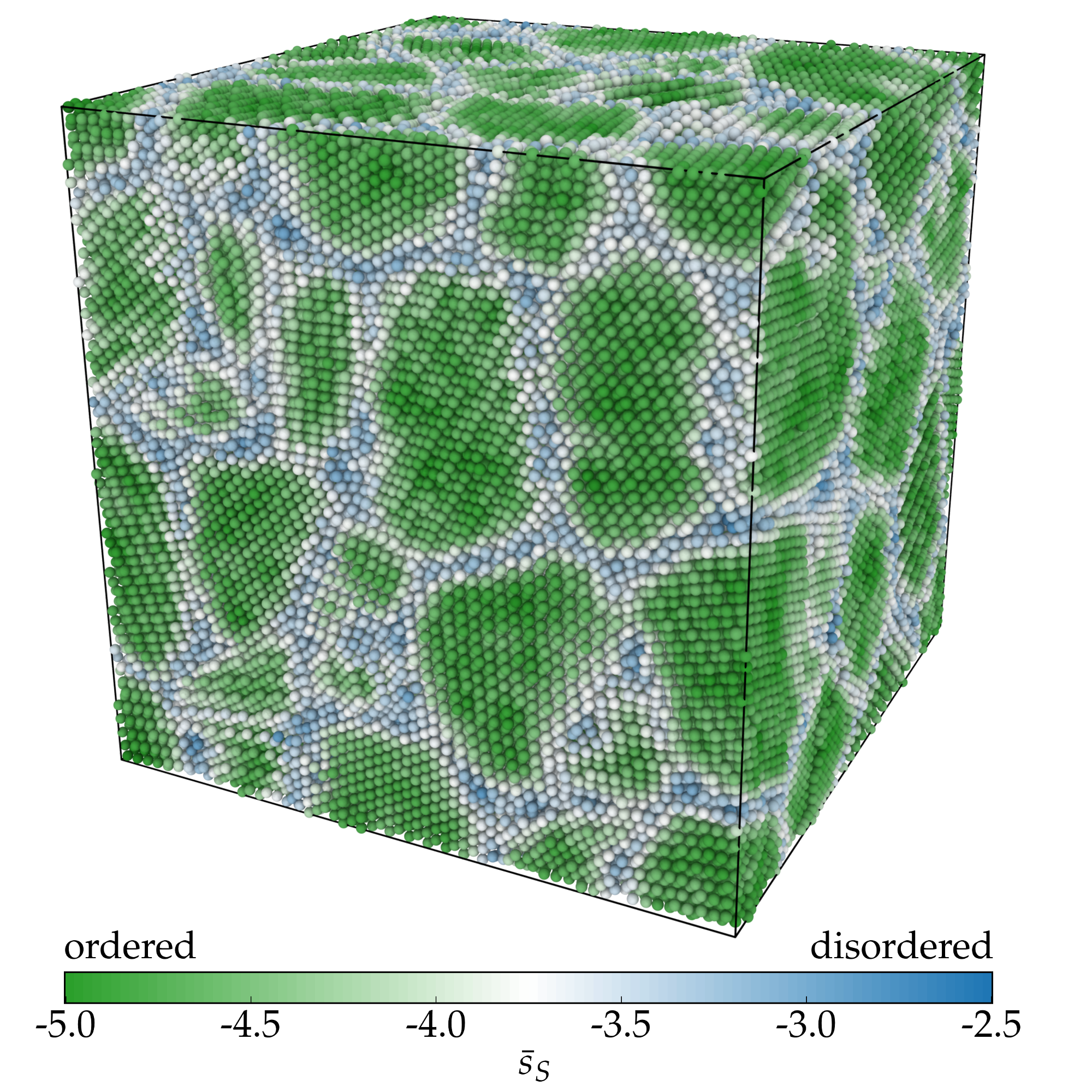}
		\caption{\label{fig:Figure3} Nanocrystalline Al with mean grain size 5 nm at 300 K.
    Atoms are colored according to $\bar{s}_S$ (see text for details).
    The colorscale is such that green and blue atoms have ordered and disordered environments, respectively.
    Image obtained with \textsc{OVITO} \cite{Stukowski09}.
    }%
	\end{center}
\end{figure}
The system is Al, as described by the potential in Ref.~\citenum{Mendelev08}.
It can be seen that the entropy fingerprint clearly brings out the nanostructure of the system and the network of grain boundaries.
This indicates that the entropy fingerprint can also work in inhomogeneous situations where different atomic environments coexist.

\section{Identification of crystal structures}

In the previous section we have shown that $\bar{s}_S$ is able to distinguish liquid-like from solid-like atomic environments.
We will now explore the possibility of distinguishing between fcc, hcp, bcc and liquid-like atomic environments.
As we shall see, this is best achieved if we accompany our definition of local entropy with a measure of local enthalpy.

The local enthalpy is easily defined if we consider an interatomic potential $U(\mathbf{R})$ that can be decomposed into energies $U_i(\mathbf{R})$ associated to individual atoms.
Here $\mathbf{R}$ denotes the atomic coordinates of an $N$ atom system.
The expression that we shall use is then,
\begin{equation}
  s_H^i = U_i(\mathbf{R})+ PV/N
  \label{eq:enthalpy_parameter}
\end{equation}
where $P$ and $V$ are the system's pressure and volume, respectively and, for simplicity, we have partitioned the volume of the system into $N$ equal parts.
A more complex partition criterion is also possible.
As done for the local entropy, we define an average local enthalpy,
\begin{equation}
  \bar{s}_H^i = \frac{\sum_j s_H^j f(r_{ij}) + s_H^i}{\sum_j f(r_{ij})+1}
  \label{eq:enthalpy_parameter_avg}
\end{equation}
where the symbols have the same meaning as in Eq. (\ref{eq:entropy_parameter_avg}).

\begin{figure*}
	\begin{center}
		\includegraphics[width=0.95\textwidth]{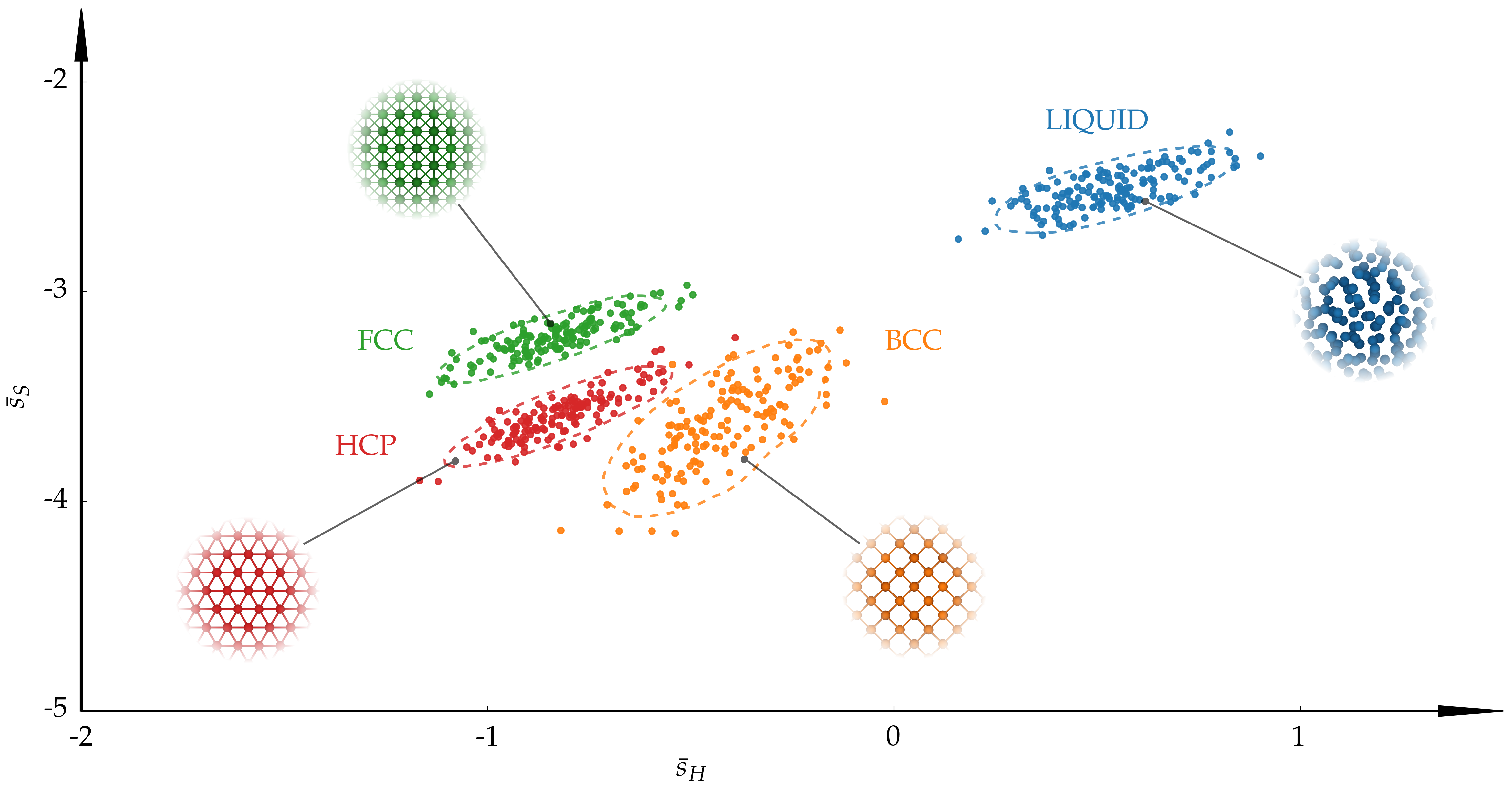}
		\caption{\label{fig:Figure4} Joint probability distributions of $\bar{s}_H$ and $\bar{s}_S$ ($P(\bar{s}_H,\bar{s}_S)$) of the fcc, hcp, bcc, and liquid phases of the Lennard-Jones system (see text for simulation details).
       The dashed lines are the iso-probability lines for a probability equal to $1/10 \: \max \{ P(\bar{s}_H,\bar{s}_S) \} $.
       The scattered points are 150 random samples of $\bar{s}_H$ and $\bar{s}_S$ over the trajectory in each phase.
       The solid atomic configurations correspond to $\{100\}$ planes of bcc and fcc crystals, and the basal plane of an hcp crystal at 0 K.
    }
	\end{center}
\end{figure*}

We calculated the joint probability distributions of $\bar{s}_H$ and $\bar{s}_S$ ($P(\bar{s}_H,\bar{s}_S)$) of the fcc, hcp, bcc, and liquid phases of the Lennard-Jones system described in Section \ref{sec:PairEntropy}.
For this purpose we simulated systems in each of those phases for 200 ps.
The thermodynamic conditions were the same as described in Section \ref{sec:PairEntropy}.
We used the following parameters to define $\bar{s}_H$ and $\bar{s}_S$: $r_{m}=r_{a}=2.5$, and $\sigma=0.1$.
The $P(\bar{s}_H,\bar{s}_S)$ of each phase are shown in Fig.~\ref{fig:Figure4}.
Each $P(\bar{s}_H,\bar{s}_S)$ was normalized to one.

We now discuss the results in Fig.~\ref{fig:Figure4}.
We first notice that the distributions of the different phases in Fig.~\ref{fig:Figure4} have minimal overlap and therefore $\bar{s}_H$ and $\bar{s}_S$ are useful fingerprints.
As in the case of Na and Al, the distributions of liquid and solid phases are very far apart and therefore the fingerprints distinguish very well between liquid-like and solid-like environments.
The distributions in the solid phases are clustered together in the region of low enthalpy and entropy, and it is easy to distinguish between the structures using $\bar{s}_H$ and $\bar{s}_S$.
We analyze in detail the challenging case of fcc and hcp.
Both fcc and hcp structures are formed by stacking of close-packed planes.
However, they differ in the way the close-packed planes are stacked.
For this reason, these structures are usually not easy to discriminate.
As seen in Fig.~\ref{fig:Figure4}, the fingerprints introduced in this work discriminate well between fcc and hcp configurations.
However, a large value of $r_a$ was necessary.

\section{Conclusions}

To conclude, the degree of success of the entropy based fingerprint is at first sight surprising.
However, the root of this success must lie on the point of view taken here that does not directly focus on the local geometry but on properties of deeper thermodynamic significance, like local entropy and enthalpy.
It also points to the usefulness of looking at old problems from a different standpoint.

\appendix

\section{Computational details}

\label{sec:appendixA}

We performed molecular dynamics (MD) simulations using \textsc{LAMMPS} \cite{Plimpton95}.
We employed an anisotropic Parrinello-Rahman barostat \cite{Parrinello81} and the stochastic velocity rescaling thermostat \cite{Bussi07}.
The fingerprints were programmed in a development version of \textsc{PLUMED~2} \cite{Tribello14}.

The Lennard-Jones simulations were performed at temperature $T=1.15$ and pressure $P=5.68$ (solid-liquid coexistence\cite{Hansen69}).
As usual, we use Lennard-Jones units \cite{FrenkelBook}, i.e. $\sigma=1$ and $\epsilon=1$.
The Lennard-Jones potential was truncated at 2.5 and tail corrections were included.
The time step for the integration of the equations of motion was 0.002.
The relaxation times of the barostat and thermostat were 5 and 0.05, respectively.

Na and Al were simulated using embedded atom models (EAM)\cite{Wilson15, Sturgeon00}.
The time step for the integration of the equations of motion was 2 fs.
For Na we set the temperature at 350 K, close to the melting temperature (366 K) of the model.
For Al the temperature was set to 900 K, near the melting temperature 931 K.
In both cases the pressure was set to its standard atmospheric value.
The relaxation times of the barostat and thermostat were 10 ps and 0.1 ps, respectively.
The results presented in Fig. \ref{fig:Figure2} were obtained by performing independent simulations in the liquid and solid phases of Na and Al at the above cited temperatures.
Each simulation had a length of 200 ps and the distributions of $s_S$ and $\bar{s}_S$ were calculated taking samples every 1 ps.

The configuration of the nanocrystalline Al was constructed using Voronoi tesselation\cite{Piaggi15, Meyers06}.
The mean grain size was 5 nm and the system contained 255064 atoms.
We performed an annealing at 600 K for 0.2 ns, then the temperature was ramped to 300 K in 0.2 ns, and finally the temperature was kept constant at 300 K for 0.2 ns.
For these simulations we employed a different EAM potential\cite{Mendelev08}.
The configuration in Fig. \ref{fig:Figure3} corresponds to the last in this trajectory.
The simulation details were the same as those used for Al above.

EAM potentials \cite{Daw84, Finnis84} have a natural way to partition the energy between the atoms as needed in Eq.~(\ref{eq:enthalpy_parameter}), i.e.
\begin{equation}
  U_i(\mathbf{R}) = \sum\limits_{j\neq i} \phi(r_{ij}) + F \left (\sum\limits_{j\neq i}\rho_{\mathrm{atom}}(r_{ij}) \right)
  \label{eq:energy_partition}
\end{equation}
where $\phi$ is a pairwise potential, $F$ is the embedding energy function, and $\rho_{\mathrm{atom}}$ is the electron charge density function.
We have used this partition criterion.

\begin{acknowledgments}
This research was supported by the NCCR MARVEL funded  by  the  Swiss  National  Science  Foundation.
The authors also acknowledge funding from the European Union Grant No. ERC-2014-AdG-670227 / VARMET.
The computational time for this work was provided by the Swiss National Supercomputing Center (CSCS) under project ID mr3.
Calculations were performed in CSCS cluster Piz Daint.
\end{acknowledgments}

\end{document}